\documentstyle{elsart}
\def\srs{\scriptscriptstyle}
\def\hacm{\hspace{-\arraycolsep}}
\def\arraylinestretch{15pt}
\def\arraylinestretchlarge{17pt}
\newcommand{\dif}[2]{\protect
\frac{\protect\partial{}#1}{\protect\partial{}#2}}
\newcommand{\im}{{\rm i}}
\newcommand{\ex}{{\rm \/e\/}}
\renewcommand{\vec}[1]{\ifmmode
\mathchoice{\mbox{\boldmath$\displaystyle\bf#1$}}
{\mbox{\boldmath$\textstyle\bf#1$}}
{\mbox{\boldmath$\scriptstyle\bf#1$}}
{\mbox{\boldmath$\scriptscriptstyle\bf#1$}}\else
{\mbox{\boldmath$\bf#1$}}\fi}
\newcommand{\vecsp}{{}}
\newcommand{\df}{{\rm d}}
\begin{document}
\begin{frontmatter}
\title{Quantile Motion and Tunneling}
\author{S. Brandt}, \author{H.D. Dahmen},
\author{E. Gjonaj\thanksref{DAAD}} and \author{T. Stroh}
\address{Fachbereich Physik, Universit\"at Siegen,
         57068 Siegen, Germany}
\thanks[DAAD]{Supported by the Deutsche Akademische Austauschdienst (DAAD)}


\begin{abstract}
The concepts of quantile position, trajectory, and velocity are 
defined. For a tunneling quantum mechanical wave packet, it is 
proved that its quantile position always stays behind that of a 
free wave packet with the same initial parameters. In quantum 
mechanics the quantile trajectories are mathematically identical to 
Bohm's trajectories. A generalization to three dimensions is given. 
\end{abstract}
\end{frontmatter}

The discussion of tunneling times has a long history in the 
literature, for a review see \cite{haug-stoev}. Recent 
developments are, e.g., the ``tunneled flux'' approach \cite{Dumont},
the operational projector approach to tunneling times \cite{Muga},
and the calculation of tunneling times in the framework of the
Dirac equation \cite{Challinor}. Among other findings it has been 
reported that the velocity of a particle can be larger within a 
repulsive barrier than outside the barrier and that instantaneous 
tunneling can occur. Great care has to be exercised, however, 
in defining a velocity in quantum mechanics, since the velocity 
definition of classical mechanics requires the concept of a 
trajectory for a point particle which breaks down in 
quantum mechanics. In this letter we introduce a definition of 
a velocity which is strictly based on probability concepts.
\section{Definitions}
\label{Definitons}
For any probability density $\varrho(x)$ the quantile $x_Q$ 
associated with the probability $Q$ is defined in the mathematical 
literature, c.f., e.g., \cite{Kendall,Brandt}, by
\[
   Q = \int_{-\infty}^{x_{Q}} \varrho (x) \,\df x \: .
\]
For a position- and time-dependent probability density $\varrho(x,t)$
we introduce the time-dependent {\em quantile position} $x_P(t)$ 
through
\begin{equation}
   P = \int_{x_{P} (t)}^{\infty} \varrho (x,t) \,\df x \: .
\label{eq:quant-def}
\end{equation}
The change to the complementary  integration interval is chosen
merely for convenience. Essential is the transfer of the quantile 
concept from a probability density in mathematical statistics to a 
space- and time-dependent distribution describing a physics problem. 
It yields for every time $t$ a well-defined point $x_{P} (t)$.
Considering $x = x_P (t)$ as a {\em quantile trajectory} we define 
the {\em quantile velocity} $v_{P} (t) = \df x_{P} / \df t$, which 
obviously depends on the chosen value of $P$. 

In the case of a conserved probability the corresponding
probability density fulfills the continuity equation of the form
\begin{equation}
  \frac{\partial \varrho}{\partial t} (x,t) +
  \frac{\partial j}{\partial x} (x,t) = 0 \: ,
  \label{eq:cont-eq}
\end{equation}
$j (x,t)$ being the corresponding probability current density
vanishing at infinity.  Equations (\ref{eq:quant-def}) and
(\ref{eq:cont-eq}) permit a direct calculation of the quantile
velocity $v_{P}(t)$. Differentiating (\ref{eq:quant-def}) with 
respect to time we immediately obtain
\begin{equation}
  v_{P} = \frac{\df x_{P}}{\df t} =
  \frac{j (x_{P},t)}{\varrho (x_{P},t)} \: .
  \label{eq:diff-quant}
\end{equation}
Equation (\ref{eq:diff-quant}) is an ordinary differential equation
for the quantile position $x_{P} (t)$, an implicit solution of
which is given by (\ref{eq:quant-def}). 

In an experiment with quantum-mechanical wave packets the quantile
velocity can be determined on a statistical basis by time-of-flight
measurements: One prepares by the same procedure $N$ single-particle
wave packets and sets a clock to zero at the moment at which the 
spatial expectation value of a wave packet leaves the source. With 
a detector placed at position $x_1$ one registers the arrival times 
$t_{1m}$ of particles for $m=1,2,\dots ,N$ and orders them such that
$t_{11} < t_{12} < \ldots$. One picks the time $t_{1n}$ which is the
largest of the smallest $n$ times and chooses $n/N = P$. The time
$t_{1n}$ is the arrival time of the quantile $x_P$ at the position
$x_1$, i.e., $x_P(t_{1n}) = x_1$. By repeating the experiment with 
a detector at $x_2$ one obtains $t_{2n}$, etc. The points
$x_P(t_{in})$  are discrete points on the quantile trajectory
$x = x_P (t)$. If $x_1$ and $x_2$ mark the beginning and the end 
of a potential barrier then $t_{2n} - t_{1n}$ is the quantile 
traversal time of the barrier.

In the case of classical systems, e.g., classical electrodynamics, 
the quantile velocity can quite naturally be interpreted as velocity 
of an electromagnetic signal. We identify the density $\varrho (x,t) 
= w(x,t)/W$ with the ratio of the electromagnetic energy density
$w$ and the total energy $W$ of the pulse. We say the signal has left 
the transmitter, if the fraction $P$ of the total energy has left the 
transmitter. It has reached the detector if the fraction $P$ has 
been absorbed by the detector. Obviously a minimum amount of energy 
(the threshold energy $W_{\rm th}$) is needed in the detector to 
register a signal. One will therefore choose $P$ such that 
$PW \ge W_{\rm th}$.
\section{Free wave packet}
\label{Free}
\begin{figure}
\vspace{12cm}
\includegraphics{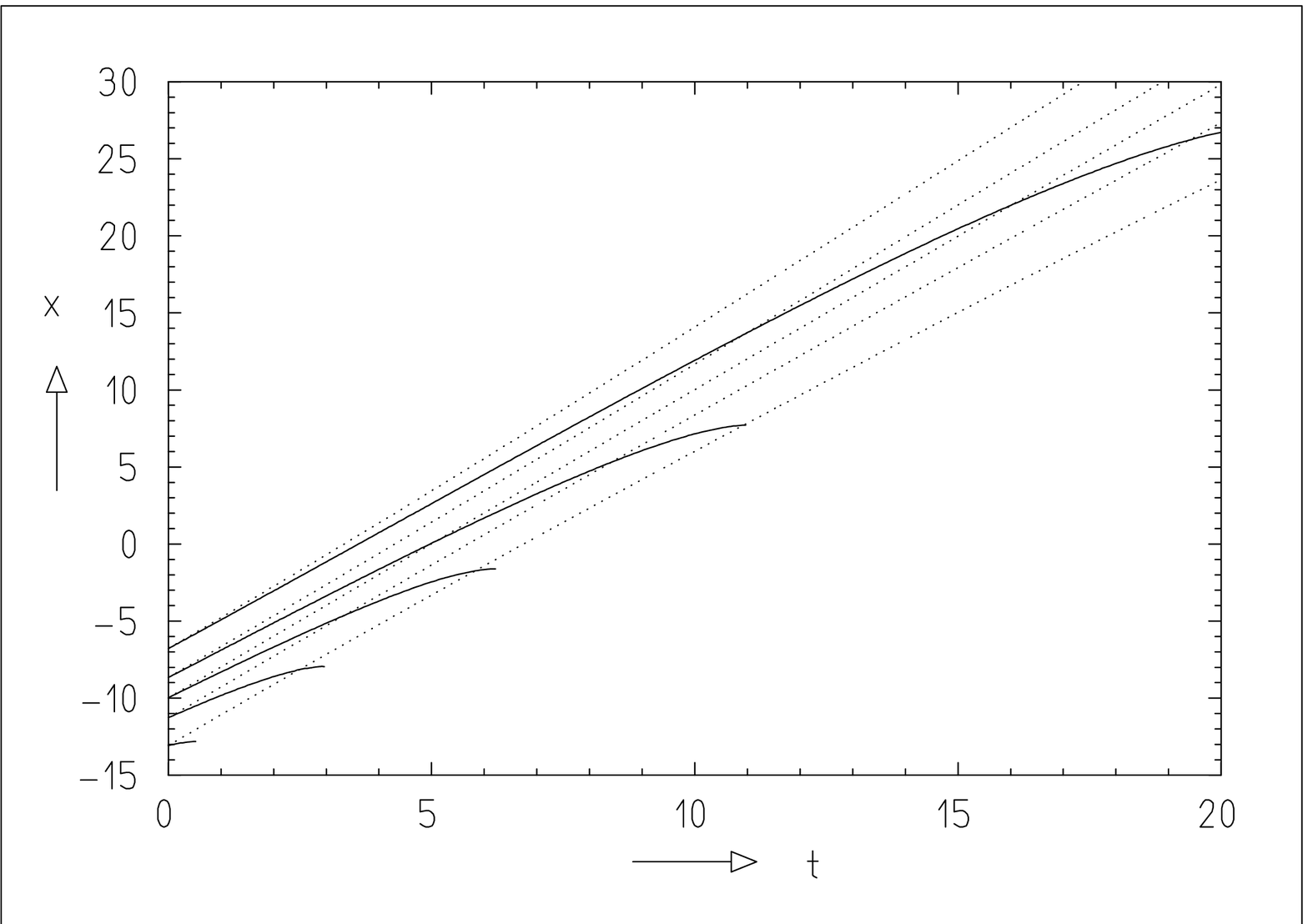}
\caption{Quantile trajectories for the wave packet (\ref{eq:Gauss-dist})
(dotted lines) and (\ref{eq:dissip-1}) (solid lines). Trajectories 
are plotted for values of $P$ between $0.1$ and $0.9$ in steps of 
$\Delta P = 0.2$. In all figures positions are given in units 
$\hbar /\protect\sqrt{{\rm eV}m}$ and times in units 
$\hbar / {\rm eV}$ with $m$ being the mass of the particle. At 
time $t = 0$ the wave packet is in both cases a Gaussian wave packet
with mean position $\bar{x} = - 10 \hbar / \protect\sqrt{{\rm eV} m}$, 
mean momentum $\bar{p} =  2 \protect\sqrt{{\rm eV}m}$, and momentum 
width $\sigma_{p} = 0.2 \protect\sqrt{{\rm eV}m}$. The wave packet
(\ref{eq:dissip-1}) is additionally characterized by a probability 
loss of temporal rate $\lambda = 0.1 \protect\hbar/{\rm eV}$.}
\label{fig:1d-quant}
\end{figure}

As an example we consider the time development of a time-dependent
Gaussian probability distribution
\catcode`@=11
\if@twocolumn
 \begin{equation}
    \begin{array}{rcl}
    \varrho (x,t) &=& \displaystyle\frac{1}{\sqrt{2 \pi} \sigma_{x}(t)}
    \exp \left\{- \frac{(x - \bar{x} -
    \bar{v} t)^{2}}{2 \sigma_{x}^{2} (t)}\right\}\: , \\ &&\\
    \sigma^{2}_{x} (t) &=& \displaystyle\sigma^{2}_{x0}
    \left(1 + \sigma^{2}_{v} t^{2} / \sigma^{2}_{x0} \right) \: .
    \end{array}
    \label{eq:Gauss-dist}
 \end{equation}
\else
 \begin{equation}
    \varrho (x,t) = \frac{1}{\sqrt{2 \pi} \sigma_{x} (t)}
    \exp \left\{- \frac{(x - \bar{x} -
    \bar{v} t)^{2}}{2 \sigma_{x}^{2} (t)}\right\}\,,
    \;\; \sigma^{2}_{x} (t) = \sigma^{2}_{x0}
    \left(1 + \frac{\sigma^{2}_{v} t^{2}}{\sigma^{2}_{x0}} \right) \, .
    \label{eq:Gauss-dist}
  \end{equation}
\fi
This probability density describes just as well (i) the marginal
distribution of a bivariate Gaussian phase-space distribution of
a classical assembly (see, e.g., \cite{br-dah-pic}) of force-free
particles of mass $m$ with initial position expectation value
$\bar{x}$, initial spatial variation $\sigma_{x0}$, and momentum
expectation value $\bar{p} = m \bar{v}$, momentum variation
$\sigma_{p} = \hbar / (2 \sigma_{x0})$, and velocity variation
$\sigma_{v}=\sigma_{p}/m$, and (ii) the spatial probability
distribution of a quantum-mechanical force-free Gaussian wave
packet with the same initial parameters as the above classical
distribution. Equation (\ref{eq:quant-def}) yields the quantile
trajectories
\[
  x_{P} (t) = \bar{x} + \bar{v} t + (\sigma_{x} (t)/\sigma_{x0})
  (x_{0} - \bar{x}) \: , \quad
\]
and thus the quantile velocities
\[
  v_{P} (t) = \bar{v} +
  \left(\sigma_{v}^{2} \, t/(\sigma_{x} (t)\sigma_{x0})\right)
  (x_{0} - \bar{x}) \: .
\]
Here $x_{0}=x_{P} (0)$ is the initial quantile position associated
with $P$ according to the condition $(1/2)\mathop{\rm erfc}
[(x_{0}-\bar{x})/(\sqrt{2}\sigma_{x0})] = P$. If we take 
(\ref{eq:Gauss-dist}) to be a classical phase-space distribution, 
we note that the quantile trajectories $x=x_P (t)$ are not 
trajectories of free particles possessing constant velocities 
$v=p/m$. Only for $P = \frac{1}{2}$, i.e.,  $x_{0} = \bar{x}$ 
is the quantile trajectory identical to a particle trajectory.  
Quantile trajectories of a free wave packet for different values 
of $P$ are shown as dotted lines in Fig.~\ref{fig:1d-quant}.
\section{Quantile motion for a non-conserved probability}
\label{Non-conserved}
Our definition (\ref{eq:quant-def}) allows a consistent description 
of motion also in the case of a non-conserved probability. We
assume that instead of a normalization to one at all times, we 
have a loss of probability,
\[
   \int_{-\infty}^{\infty}\varrho (x,t) \, \df x 
   = F(t) \leq 1 \: , \quad F(0) = 1 \: .
\]
As an illustration we consider again the time development of 
a wave packet prepared at $t=0$ to be a Gaussian distribution 
with the same parameters as in (\ref{eq:Gauss-dist}). In addition 
we introduce a temporal exponential decrease of the probability 
to find the particle anywhere in space, as for example caused by 
the presence of a constant purely imaginary potential. Thus the 
probability density may be written in the form
\begin{equation}
    \varrho_\lambda (x,t) = \frac{1}{\sqrt{2 \pi} 
    \sigma_{x} (t)} \exp \left\{- \frac{(x - \bar{x} -
    \bar{v} t)^{2}}{2 \sigma_{x}^{2} (t)}\right\}
    \exp \left\{- \lambda t \right\} \: ,
\label{eq:dissip-1}
\end{equation}
where $\lambda$ is the temporal rate of probability loss and
$\sigma_{x}(t)$ is given in equation (\ref{eq:Gauss-dist}). 
Instead of the continuity equation (\ref{eq:cont-eq}) we have now
\begin{equation}
  \frac{\partial \varrho_\lambda}{\partial t} (x,t) +
  \frac{\partial j_\lambda}{\partial x} (x,t) = - l (x,t) \: ,
\label{eq:dissip-2}
\end{equation}
$j_\lambda (x,t)$ being the current density of the non-conserved 
probability and $l (x,t) = \lambda \varrho_\lambda (x,t) \geq 0$ 
describing the density of the temporal loss rate. For the quantile 
position $x_P(t)$ we obtain the integro-differential equation
\begin{equation}
  \frac{\df x_{P}(t)}{\df t} =
  \frac{j_\lambda (x_{P},t)}{\varrho_\lambda (x_{P},t)} -
  \frac{1}{\varrho_\lambda (x_{P},t)}
  \int_{x_P(t)}^{\infty} l(x',t)\,\df x' \: .
  \label{eq:dissip-3}
\end{equation}
The solid lines in Fig.~\ref{fig:1d-quant} are quantile 
trajectories  for different values of $P$. The curves end at 
the quantile position $x_{P\rm max}$ where the integral over 
the loss rate on the right--hand side of (\ref{eq:dissip-3}) 
equals the probability current density $j_\lambda (x,t)$. 
At later times the probability loss term in (\ref{eq:dissip-3}) 
dominates over the probability current density $j_\lambda (x,t)$. 
For positions at values $x > x_{P\rm max}$ the condition 
(\ref{eq:quant-def}) can no longer be satisfied. Note that the 
formulation of quantile motion for a non-conserved probability 
has general validity, since Eqs.~(\ref{eq:dissip-2}) and 
(\ref{eq:dissip-3}) do not depend on the details of the 
probability loss.
\section{Tunneling}
\label{Tunneling}
In the following we show that for the same value of $P$ and 
for two identically prepared, tunneling and free wave packets, 
the quantile position of the tunneling wave packet in the 
transmission region of the potential remains at any given time 
$t$ behind the quantile position of the free wave packet, i.e.,
the arrival time of the tunneled quantile behind the barrier 
is always later than that of the free wave packet. We consider 
the tunneling of a wave packet with spectral function 
$\widetilde{\psi}(k)$ in momentum space through a repulsive 
square potential barrier, $V(x)=V\geq 0$ for $|x|<a$ and 
$V(x)=0$ for $|x|>a$. The probability densities of the 
tunneling and of the free wave packet with the same spectral 
function $\widetilde{\psi}(k)$ are denoted by $\varrho_{\rm T}(x,t)$ 
and $\varrho_{\rm F}(x,t)$, respectively. The functions 
\[
    P_{\rm T}(x,t) = \int_{x}^{\infty}\varrho_{\rm T}(x',t)\, \df x' \: ,
    \qquad
    P_{\rm F}(x,t) = \int_{x}^{\infty}\varrho_{\rm F}(x',t)\, \df x' 
\]
give the probabilities at time $t$ for the tunneling and the 
free particle to be found to the right of the position $x$ in 
the transmission region $x>a$ of the potential. Since 
$P_{\rm T}(x,t)$ and $P_{\rm F}(x,t)$ decrease monotonically
with $x$, our statement is always true if the condition
\begin{equation}
    \Delta P(x,t)=P_{\rm F}(x,t)-P_{\rm T}(x,t) \geq 0 
\label{eq-2}
\end{equation}
for $x>a$ is fulfilled. We consider a wave packet described by a
spectral function $\widetilde{\psi}(k)$ with a positive wave number
spectrum, i.e., $\widetilde{\psi}(k)=0$ for $k<0$. The difference 
$\Delta P(x,t)$ in the region $x>a$ is
\[
   \Delta P(x,t)=\frac{\im}{2\pi}\int_{0}^{\infty}
   \int_{0}^{\infty}[1-T(k')T^*(k'')]
   \frac{\widetilde{\psi}(k';x,t)\widetilde{\psi}^*(k'';x,t)}{
   k'-k'' + \im\varepsilon} \, \df k'\df k'' \: ,
\]
where $\widetilde{\psi}(k;x,t)=\widetilde{\psi}(k)
\exp[\im  k(x-\bar{x})-\im\hbar k^2 t/(2m)]$ and
\[
   T(k) = \frac{4k\gamma}{{\cal D}(k)} \: , \qquad
   {\cal D}(k) = (k+\gamma)^2\ex^{2\im a(k-\gamma)} -
   (k-\gamma)^2\ex^{2\im a(k+\gamma)} \: ,
\]
is the transmission amplitude in momentum space for the repulsive
square potential barrier ($\gamma=(k^2-2mV/\hbar^2)^{1/2}$). We
consider $\Delta P(x,t)$ as the limiting function of a series of
parameterized functions
\[
\begin{array}{rcl}
   \Delta P_{\lambda}(x,t)\hacm&\hacm=\hacm&\hacm\displaystyle
   \frac{\im}{2\pi}\int_{0}^{\infty}
   \int_{0}^{\infty}
   \frac{{\cal D}_{\lambda}(k'){\cal D}_{\lambda}^*(k'')-
   16k'k''\gamma'\gamma''^*}{{\cal D}(k'){\cal D}^*(k'')}
   \\[\arraylinestretch]&\times&\displaystyle\hacm\frac{
   \widetilde{\psi}(k';x,t)\widetilde{\psi}^*(k'';x,t)}{k'-k''
   + \im\varepsilon} \, \df k'\df k''
\end{array}
\]
for $\lambda\to 1$, where ${\cal D}_{\lambda}(k) = 
(k+\gamma)^2\ex^{2\im a\lambda(k-\gamma)} - (k-\gamma)^2
\ex^{2\im a\lambda(k+\gamma)}$. Thus, we have 
$\Delta P_{\lambda}(x,t)=0$ for $\lambda=0$, 
so that the difference of probabilities is
\begin{equation}
   \Delta P(x,t)=\int_{0}^{\srs 1} \dif{}{\lambda}
   \Delta P_{\lambda}(x,t) \, \df \lambda \: .
\label{eq-8}
\end{equation}
Explicitly calculating the derivative in (\ref{eq-8}) yields
\begin{equation}
\begin{array}{rcl}
   \Delta P(x,t)\hacm&\hacm=\hacm&\hacm\displaystyle
   \frac{2a}{\pi}\left(\frac{4mV}{\hbar^2}\right)\left\{
   \int_{0}^{\srs 1}\left|\, \int_{0}^{\infty}
   \frac{{\cal Z}_{\lambda}(k)}{{\cal D}(k)}
   \widetilde{\psi}(k;x,t)\,\df k\, \right|^2\hacm\df\lambda\right.
   \\[\arraylinestretchlarge]
   &+&\hacm\displaystyle 4\left(\frac{4mV}{\hbar^2}\right)
   \int_{0}^{\srs 1}\left|\, \int_{0}^{\infty}
   \frac{\ex^{2\im a\lambda k}\sin(2a\lambda\gamma)}{{\cal D}(k)}
   \widetilde{\psi}(k;x,t)\,\df k\,\right|^2\hacm\df\lambda
   \\[\arraylinestretchlarge]
   &+&\hacm\displaystyle\left.\left(\frac{4mV}{a\hbar^2}\right)
   \int_{x}^{\infty}\left|\, \int_{0}^{\infty}
   \frac{\ex^{2\im ak}\sin(2a\gamma)}{{\cal D}(k)}
   \widetilde{\psi}(k;x',t)\,\df k\,\right|^2\hacm\df x'\right\} \: ,
\end{array}
\label{eq-9}
\end{equation}
where ${\cal Z}_{\lambda}(k)=(k+\gamma)\ex^{2\im a
\lambda(k-\gamma)} - (k-\gamma)\ex^{2\im a\lambda(k+\gamma)}$. 
The positivity of (\ref{eq-9}) proves the condition (\ref{eq-2}) 
and therefore the initial statement on the retardation of the 
quantile trajectory of the tunneled wave packet in the transmission 
region of the potential. This procedure can be extended to the case 
of a general non-negative potential $V(x)$ of finite range, 
introducing a segmentation of the potential into thin square 
potential barriers and then similarly parameterizing the difference 
$\Delta P(x,t)$ to obtain a positive-definite expression \cite{Gjonaj}.

\begin{figure}
\vspace{15.5cm}
\includegraphics{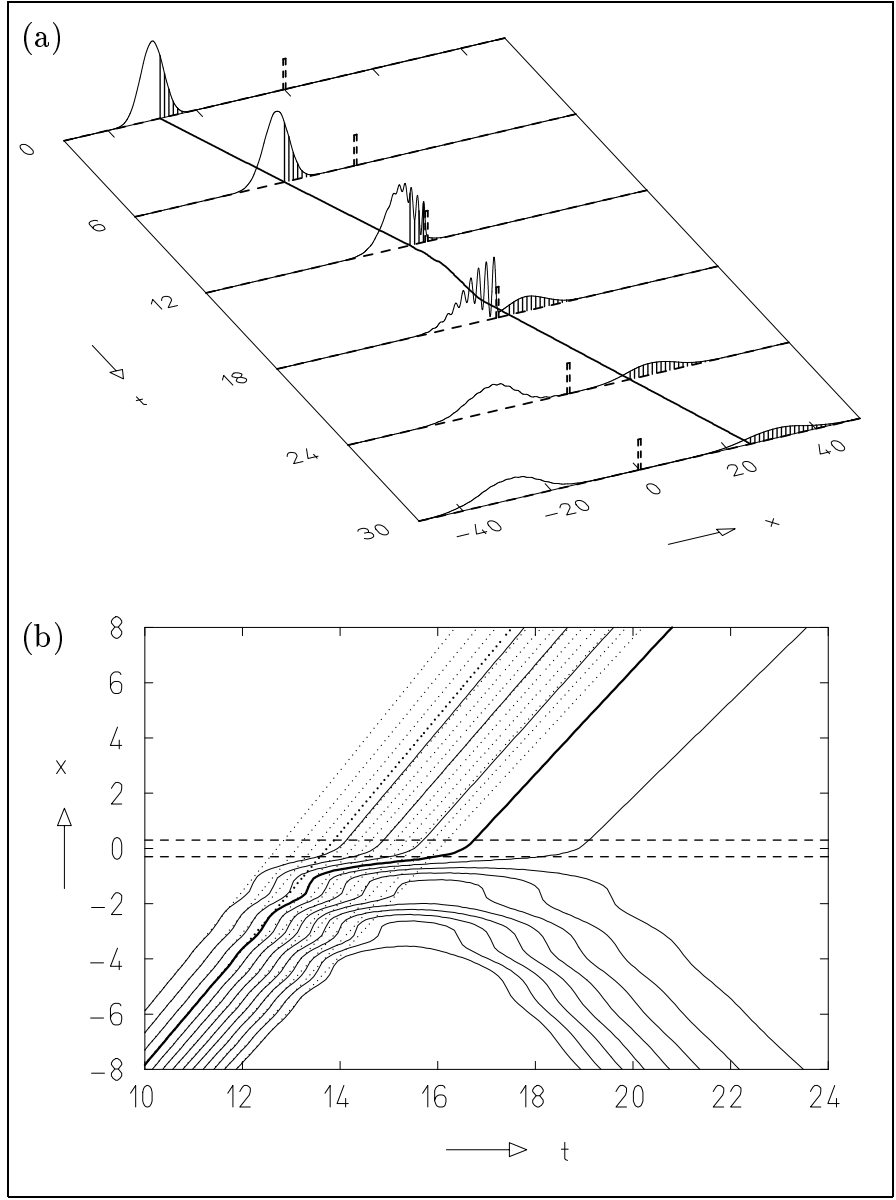}
\caption[]{(a) Time development of a Gaussian wave packet with the
same initial parameters as in Fig.~\ref{fig:1d-quant} incident
onto a repulsive potential barrier (indicated as broken line) of
height $V_{0} = 10\, {\rm eV}$ extending from $x = - 0.3
\hbar/\protect\sqrt{{\rm eV}m}$ to $x = 0.3
\hbar/\protect\sqrt{{\rm eV}m}$, where $m$ is the particle mass.
The individual graphs show the probability density $\varrho(x,t)$
as function of $x$ for fixed values of $t$. The hatched area 
comprises the probability $P = 0.25$. The thick line is the quantile
trajectory for that value of $P$. (b) Quantile trajectories $x =
x_{P} (t)$ (solid lines) for values between $P = 0.1$ and 
$P = 0.7$ in steps of $0.05$ for the wave packet shown in (a) but 
presented in a smaller part of the $x,t$ plane. The thicker line
corresponds to the trajectory shown in (a). The barrier region is
bounded by broken lines. For comparison the quantile trajectories
$x = x_{P}^{(\rm free)} (t)$ of a free wave packet are also shown 
as dotted lines. Note that $x_{P} (t) \leq x_{P}^{\rm (free)} (t)$ 
for every pair of values $P$ and $t$.}
\label{fig:2d-quant}
\end{figure}

In Fig.~\ref{fig:2d-quant}a the time development of a wave
packet incident on a barrier is shown. The packet is partly
reflected and partly transmitted. In Fig.~\ref{fig:2d-quant}b 
the quantile trajectories $x = x_{P}(t)$ for the same problem are 
presented.  For small values of $P$ the trajectories penetrate 
the barrier, for larger values they reverse their direction. We 
observe that in all cases the  quantile velocity within the barrier 
region is smaller in absolute value than far from the barrier.
\section{Relation to Bohmian mechanics}
\label{Bohm}
Bohm's interpretation of quantum mechanics \cite{bohm52} introduces 
particle trajectories $x = x(t)$ satisfying the equation of motion
$\df x(t)/\df t = j(x(t),t)/\varrho(x(t),t)$. For the solution
of this differential equation the initial condition $x(0) = x_0$
is needed as a hidden parameter. If we set Bohm's initial position 
$x_0$ equal to the initial quantile position $x_{P}(0)$ the
quantile trajectories are mathematically identical to Bohm's 
particle trajectories. Conceptually, however, they are based on 
the probability interpretation of the conventional quantum mechanics.

Leavens and Aers \cite{leav90} and Spiller et~al.\ 
\cite{spill-et-al90} realized that Bohm's interpretation of 
quantum mechanics --- reintroducing the point-particle concept --- 
offers the opportunity of unambiguously defining the velocity as 
in classical mechanics (see also \cite{Berndl,Daumer}). In contrast 
to their work our approach is based on the conventional probabilistic 
concepts and does not depend on the interpretation given by Bohm. 
In Ref.~\cite{leav90} tunneling times are given which are based 
on the concept of Bohm trajectories.  McKinnon and Leavens 
\cite{mck-leav} pointed out the significance with respect to tunnel 
times of a particular Bohm trajectory for $P=|T|^2$, with $|T|^2$ 
being the transmission probability. Wu and Sprung \cite{Wu} have 
discussed quantum probability patterns for stationary problems. 

\section{Generalization to three dimensions}
\label{Generalization}
\begin{figure}
\vspace{16cm}
\includegraphics{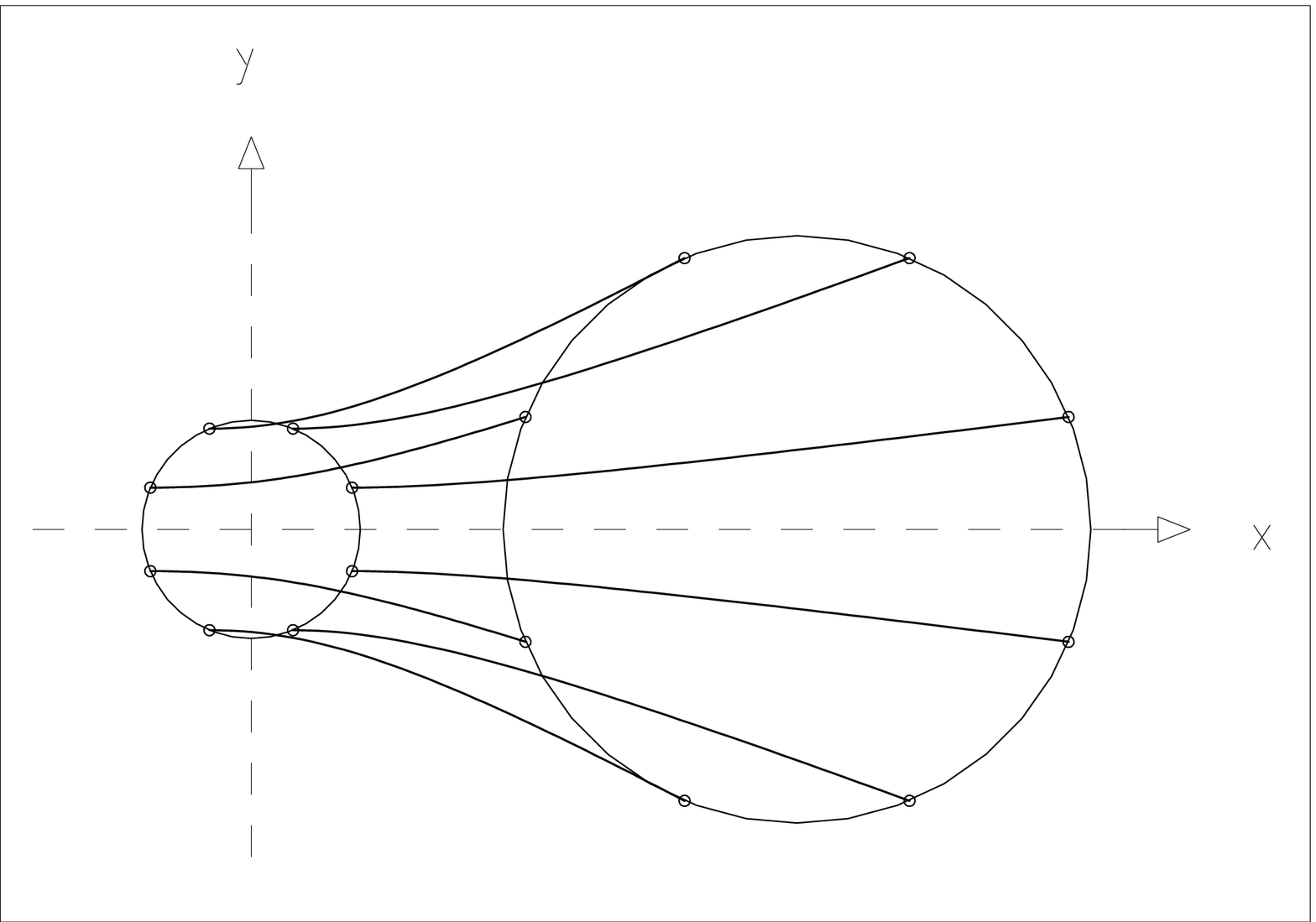}
\caption{The expectation value of a free three-dimensional 
spherically symmetric Gaussian wave packet, which initially 
(at time $t=0$) is at the origin, moves in the positive $x$ 
direction. As initial volume $V_0$ comprising the probability 
$P$ a sphere around the origin is chosen.  Quantile trajectories 
$\protect\vec{x}_P(t)$ of points which at $t=0$ lie on the surface 
of $V_0$ at later times lie on the surface of a volume $V_t$ which 
also comprises the same probability $P$. In this simple example all 
volumes $V_t$ are spheres. The plot shows the cuts $z=0$ through two 
spheres $V_0, V_t$ which are circles and trajectories in the $x,y$ plane.}
\label{fig:3d-quant}
\end{figure}

The equivalence of equations (\ref{eq:quant-def}) and 
(\ref{eq:diff-quant}) has been noticed before, see, e.g., 
\cite{Berndl,leav93}. In \cite{Berndl} it was considered as an 
artifact of low dimensionality. In both papers, this equivalence was 
not related to the probability interpretation made possible by equation 
(\ref{eq:quant-def}). However, the generalization to three dimensions 
is straightforward, if one considers that the interval between two 
quantile positions $x_{P_1}(t)$ and $x_{P_2}(t)$ always contains the 
same amount of probability $Q = P_1 - P_2$ and that all trajectories 
with initial positions $x_{P_1}(0) < x_P(0) < x_{P_2} (0)$ stay in 
the interval $x_{P_1} (t) < x_P(t) < x_{P_2} (t)$. This suggests that 
in three dimensions trajectories defined by the equation of motion
\begin{equation}
  \frac{\partial \vec{x}_P(t, \vec{x})}{\partial t} =
  \vec{v} (\vec{x}_P,t)
   = \frac{\vec{j}(\vec{x}_P,t)}{\varrho (\vec{x}_P,t)}
  \label{eq:3d-diff-traj}
\end{equation}
and having initial positions $\vec{x}_P (0, \vec{x}) = \vec{x}$ within 
a volume $V$ containing an amount $P$ of probability stay within a 
volume $V_t$ containing the same amount of probability. To prove 
this statement we introduce the time-dependent base of tangent vectors 
$\vec{\eta}\vecsp^i (t,\vec{x}) = \partial \vec{x}_P(t, \vec{x})
/\partial x_i$, where $i = 1,2,3$. Their Jacobian is given by the
expression $D(t,\vec{x}) = \vec{\eta}\vecsp^1 (t, \vec{x}) \cdot
[ \vec{\eta}\vecsp^2 (t,\vec{x}) \times \vec{\eta}\vecsp^3 (t,\vec{x})]$,
whereas its time derivative satisfies
\[
  \frac{\partial D}{\partial t} (t, \vec{x}) = D(t, \vec{x})
  \vec{\nabla}_{\vec{x}_P} \cdot \vec{v} (\vec{x}_P (t, \vec{x}),t) \: .
\]
In particular at the initial time $t = 0$ we have $D (0) = 1$. The total
time derivative of the density is then given by
\[
  \frac{\df\varrho (\vec{x}_P,t)}{\df t} = - \varrho (\vec{x}_P,t)
  \vec{\nabla}_{\vec{x}_P} \cdot \vec{v} (\vec{x}_P, t) \: ,
\]
so that we find the product $p(t,\vec{x}) = \varrho (\vec{x}_P
(t,\vec{x}), t) D(t,\vec{x})$ to be time independent,
\[
  \frac{\partial p}{\partial t} (t, \vec{x}) = 0 \: .
\]
This implies that an integral over a finite volume $V_0$,
\[
  \int_{V_0} p (t, \vec{x}\vecsp') \,\df^3 \vec{x}\vecsp' = P \: ,
\]
is time independent. An amount of probability
\[
  P = \int_{V_0} p (0, \vec{x}\vecsp') \,\df^3 \vec{x}\vecsp' =
  \int_{V_0} \varrho (\vec{x}, 0) \,\df^3 \vec{x}
\]
contained initially in a volume $V_0$ is at time $t$ given by
\[
  P = \int_{V_0} \varrho (\vec{x}_P (t, \vec{x}\vecsp'), t)
  D(t, \vec{x}\vecsp') \,\df^3 \vec{x}\vecsp' \: .
\]
The time-dependent substitution $\vec{x} = \vec{x}_P (t,
\vec{x}\vecsp')$ yields
\[
  P = \int_{V_t} \varrho (\vec{x},t) \,\df^3 \vec{x}
\]
with the volume $V_t$ given by all the points $\vec{x}_P(t,\vec{x})$
at time $t$ lying on trajectories with initial positions $\vec{x}_P
(0,\vec{x}) = \vec{x}$ in the volume $V_0$. For a simple example
quantile trajectories of a three-dimensional wave packet are shown
in Fig.~\ref{fig:3d-quant}. Requiring the quantity $P$ to be 
time independent, is a necessary condition for the validity of 
(\ref{eq:3d-diff-traj}). Sufficient conditions will be discussed
in a forthcoming publication. We note that a corresponding statement
holds also for a system with a loss term in the continuity equation,
c.f.\ (\ref{eq:dissip-2}).
\section{Concluding remarks}
\label{Concluding}
We wish again to emphasize that the concept of quantile velocities 
is not limited to problems of classical statistical assemblies and 
problems of quantum mechanics but can easily be extended to describe 
the propagation of pulses of electromagnetic energy in vacuum, in 
dispersive and absorptive media, and in wave guides \cite{Gjonaj}.
Trajectories derived from (\ref{eq:3d-diff-traj}) for electromagnetic
phenomena (with $\vec{j}$ and $\varrho$ being the energy current
density and the energy density, respectively) were already discussed
by Holland \cite{holland93} but no quantile interpretation was given.

\newpage
\newpage
\newpage


\begin{thebibliography}{9}
\bibitem{haug-stoev} 
		 E.H. Hauge and J.A. St\o vneng, 
		 Rev. Mod. Phys. 61 (1989) 917. \\
		 C.R. Leavens, {\em Tunneling and its Implications},
		 edited by D. Mugnai, A. Ranfagni and L.S. Schulman 
		 (World Scientific, Singapore, 1997).
\bibitem{Dumont} R.S. Dumont, T.L. Marchioro, Phys. Rev. A 47 (1993) 85.
\bibitem{Muga}   S. Brouard, R. Sala, J.G. Muga, Phys. Rev. A 47 
		 (1994) 4312.\\
		 J.G. Muga, S. Brouard, D. Mac\'{\i}as, Ann. of Phys. (NY)
		 240 (1995) 351. \\
		 J.G. Muga, V. Delgado, R.F. Snider, Phys. Rev. B 52
		 (1995) 16381.
\bibitem{Challinor} A. Challinor, A. Lasenby, S. Somaroo,
		   C. Doran, S. Gull, Phys. Lett. A 227 (1997) 143.
\bibitem{Kendall} M.G. Kendall, A. Stuart, {\em The Advanced Theory of
                 Statistics}, Vol. 1, 3.~ed., Charles Griffin, London 1969.
\bibitem{Brandt} S. Brandt, {\em Statistical and Computational Methods in
                 Data Analysis}, 2.~ed., North Holland, Amsterdam, 1976.
\bibitem{br-dah-pic} S. Brandt, H.D. Dahmen, {\em The Picture Book of
                 Quantum Mechanics}, 2.~ed., Springer-Verlag, New York, 1995.
\bibitem{Gjonaj} E. Gjonaj, Ph.D. Thesis, Universit\"at Siegen, 1998.
\bibitem{bohm52} D. Bohm, Phys. Rev. 85 (1952) 166; 85 (1952) 180.\\
                 D. Bohm, B.J. Hiley and P.N. Kaloyerou, Phys. Rep. 144
                 No. 6 (1987) 321-375.
\bibitem{leav90} C.R. Leavens, Solid State Commun. 76 (1990) 253.\\
                 C.R. Leavens and G.C. Aers, Solid State Commun. 78 (1991) 1015.
\bibitem{spill-et-al90} T.P. Spiller, T.D. Clark, R.J. Prance,
                 H. Prance, Europhys. Lett. 12 (1990) 1.
\bibitem{Berndl} K. Berndl, {\em Bohmian Mechanics and Quantum Theory. An
                 Appraisal}, edited by J.T. Cushing, A. Fine and
                 S. Goldstein (Kluwer, 1996).
\bibitem{Daumer} M. Daumer, {\em Bohmian Mechanics and Quantum Theory. An
                 Appraisal}, edited by J.T. Cushing, A. Fine and
                 S. Goldstein (Kluwer, 1996).
\bibitem{mck-leav} W.R. McKinnon and C.R. Leavens, Phys. Rev. A 51
                 (1995) 2748.
\bibitem{Wu}     H. Wu and D.W.L. Sprung, Phys. Lett. A 183 (1993) 413-417.
\bibitem{leav93} C.R. Leavens, Phys. Lett. A 178 (1993) 27.
\bibitem{holland93} P.R. Holland, Phys. Rep. 224 (1993) 95.
\end{thebibliography}
\end{document}